\documentclass[aps,prev,twocolumn,preprintnumbers,floatfix,nofootinbib]{revtex4-1}
\usepackage{amsmath, amsthm}
\usepackage[dvips]{graphicx}
\usepackage{graphicx}
\usepackage{bm}
\usepackage{times}

\usepackage{slashed}
\usepackage{color}

\usepackage{color}
\newcommand{\lsim}{\lower0.6ex\vbox{\hbox{$ \buildrel{\textstyle <}\over{\sim}\ $}}}
\newcommand{\gsim}{\lower0.6ex\vbox{\hbox{$ \buildrel{\textstyle >}\over{\sim}\ $}}}
\newcommand{\beq}{\begin{equation}}
\newcommand{\eeq}{\end{equation}}

\newcommand{\mnras}{Mon. Not. R. Astron. Soc.}

\newcommand{\aap}{Astron. Astrophys.}

\newcommand{\apjs}{Astrophys. J. Suppl. Ser.}

\def\ls{\mathrel{\lower4pt\vbox{\lineskip=0pt\baselineskip=0pt
           \hbox{$<$}\hbox{$\sim$}}}}
\def\gs{\mathrel{\lower4pt\vbox{\lineskip=0pt\baselineskip=0pt
           \hbox{$>$}\hbox{$\sim$}}}}

\newcommand{\be}{\begin{equation}}
\newcommand{\ee}{\end{equation}}
\newcommand{\bea}{\begin{eqnarray}}
\newcommand{\eea}{\end{eqnarray}}

\begin{document}

\begin{flushright} MIFPA-14-34\end{flushright}
\title{Dark Matter from Late Invisible Decays to/of Gravitinos}

\author{Rouzbeh Allahverdi$^{a}$}
\author{Bhaskar Dutta$^{b}$}
\author{Farinaldo S. Queiroz$^{c}$}
\author{Louis E. Strigari$^{b}$}
\author{Mei-Yu Wang$^{b}$}

\affiliation{$^{a}$Department of Physics and Astronomy, University of New Mexico, Albuquerque, NM 87131\\
$^{b}$ Department of Physics and Astronomy, Mitchell Institute for Fundamental Physics and Astronomy, Texas A \& M University, College Station, TX 77843-4242.\\
$^{c}$ Department of Physics and Santa Cruz Institute for Particle Physics
University of California, Santa Cruz, CA 95064, USA
}

\pacs{}
\date{\today}
\vspace{1cm}

\begin{abstract}
In this work, we sift a simple supersymmetric framework of late invisible decays to/of the gravitino. We investigate two cases where the gravitino is the lightest supersymmetric particle or the next-to-lightest supersymmetric particle.
The next-to-lightest supersymmetric particle decays into two dark matter candidates and has a long lifetime due to gravitationally suppressed interactions. 
However, because of the absence of any hadronic or electromagnetic products, it satisfies the tight bounds set by big bang nucleosynthesis and cosmic microwaved background. One or both of the dark matter candidates produced in invisible decays can contribute to the amount of dark radiation and suppress perturbations at scales that are being probed by the galaxy power spectrum and the Lyman-alpha forest data.    
We show that these constraints are satisfied in large regions of the parameter space and, as a result, the late invisible decays to/of the gravitino can be responsible for the entire dark matter relic abundance. 
\end{abstract}

\maketitle

\section{INTRODUCTION}
\label{introduction}
There are various lines of evidence for the existence of dark matter (DM) in the universe~\cite{dm}, but its identity is still unknown and remains one the most important problems at the interface of cosmology and particle physics. Given the many ongoing direct and indirect DM detection experiments, along with collider searches trying to pin down the nature of DM, this puzzle is expected to be solved in the foreseeable future.

Supersymmetric (SUSY) extensions of the standard model (SM) with $R$-parity conservation provide a natural candidate for DM. In these models, the lightest supersymmetric particle (LSP) is stable and hence can account for the DM particle. In the minimal supersymmetric standard model (MSSM), the LSP is either the lightest neutralino ${\tilde \chi}^0_1$ or the gravitino ${\tilde G}$. 

The presence of the gravitino results in important cosmological constraints on SUSY models. If gravitino is not the LSP, then it will decay to the LSP and its SUSY partner. If gravitino is the LSP, then the next-to-lighttest supersymmetric particle (NLSP) will decay to the gravitino. Due to Planck suppressed interactions of ${\tilde G}$ with other particles~\cite{moroithesis}, these decays have long lifetimes. In particular, if $m_{\tilde G} < 40$ TeV, the decay to/of the gravitino will occur after the onset of the big bang nucleosynthesis (BBN). Such decays are tightly constrained by cosmological considerations from BBN and comic microwave background (CMB). 
Late decays of neutral or charged particles through electromagnetic and hadronic channels are severely restricted by BBN constraints~\cite{moroibbn,olivebbn}.
Moreover, late decays that release energy in the electromagnetic mode can give rise to a chemical potential for the cosmic microwave background (CMB) photons~\cite{subircmb,hucmb}, which is constrained by observations~\cite{cobecmb}. 
Late injection of energetic neutrinos, which can produce secondary particles, is constrained by BBN bounds as well as CMB limits on the amount of extra radiation and structure formation~\cite{moroinu}. These bounds severely constrain the abundance of the NLSP and, through that, put tight limits on the reheating of the universe. These studies in the context of axion/axino production from invisible decays have been discussed in Ref.\cite{Queiroz:2014ara}. For other recent dark radiation setups see Ref.\cite{Allahverdi:2014ppa}.


In this paper, we investigate a SUSY scenario that can accommodate invisible decay to/of the gravitino. The model is a minimal extension of the MSSM and has two DM candidates. 
One of the DM candidates is the LSP and the other one is an $R$-parity even fermion $N$ with ${\cal O}({\rm GeV})$ mass that is a singlet under the SM gauge group. The LSP in this model can be either the gravitino or the SUSY partner of $N$ (denoted by $\tilde N$).
The model is well motivated due to its ability to generate the baryon abundance of the universe at temperatures well below the electroweak scale, and to explain the apparent coincidence between the the observed DM and baryon energy densities 
~\cite{Bhaskar1,Rabi3}. The invisible decays involving the gravitino are are ${\tilde N} \rightarrow {\tilde G} + N$ and ${\tilde G} \rightarrow {\tilde N} + N$ that respectively take place for a graviitno NLSP and ${\tilde N}$ NLSP and vice versa. 
Although both decays involve gravitationally suppressed interactions, and hence have a long lifetime, they circumvent the severe BBN and CMB constraints since they do not include electromagnetic or hadronic products.  

However, depending on the mass ratio between ${\tilde N}$ and ${\tilde G}$, it is possible that one or both of the decay products 
are relativistic during the epoch of matter-radiation equality. 
They may then contribute to the amount of dark radiation 
and suppress DM perturbations scales that are probed by the galaxy clustering and the Lyman-alpha forest data. We will show that for the current data
these constraints are satisfied 
in large regions of the parameter space. As a result, the late decays can be responsible for the entire DM relic abundance in this model. Moreover, in a broader context, late invisible decays to/of the gravitino considerably relax the constraints on reheating of the universe in SUSY models~\cite{reheating}.  

The structure of this paper is as follows. In Section~\ref{sec:model} we discuss the model. In Section~\ref{sec:invisible}, we discuss the late invisible decays that involve the gravitino. We discuss production of dark radiation in Section~\ref{sec:derive}, and constraints from structure formation in Section~\ref{sec:structure}. We present our results in Section~\ref{sec:result}. Finally, we close this paper by concluding it in Section~\ref{sec:conclusion}.

\section{THE MODEL}
\label{sec:model}
The model is an 
extension of the MSSM that contains iso-singlet color-triplet superfields $X$ and ${\bar X}$ with respective hypercharges $+4/3$ and $-4/3$, and a singlet superfield $N$. The superpotential of this model is given by~\cite{Rabi2}
\bea \label{superpot}
W & = & W_{\rm MSSM} + W_{\rm new} \, , \nonumber \\
W_{\rm new} & = & \lambda_{i} X N u^c_{i} + \lambda^\prime_{i j} {\bar X} {d^c_i} d^c_j + M_X X {\bar X} + \frac{M_N}{2} N N \, . \nonumber \\
& & \,
\eea
Here $i,~j$ denote flavor indices (color indices are omitted for simplicity), with $\lambda^\prime_{i j}$ being antisymmetric under $i \leftrightarrow j$. We assign quantum number $+1$ under $R$-parity to the scalar components of $X,~{\bar X}$ and the fermionic components of $N$.
As shown in~\cite{Rabi2,Bhaskar1,Rabi3}, with two (or more) copies of $X,~{\bar X}$ one can generate the baryon asymmetry of the universe from the interference of tree-level and one-loop diagrams in decay processes governed by the $X,~{\bar X}$ interactions.

The exchange of $X,\bar{X}$ particles in combination with the Majorana mass of $N$ lead to 
double proton decay $p p \to K^+ K^+$. Current limits on this process from the Super-Kamiokande experiment~\cite{miura} require that
$\vert \lambda_1 \lambda^{\prime}_{12} \vert^2 \leq 10^{-10}$ for $M_N \sim 100$ GeV. This is also enough to satisfy constraints from $K^0_s-{\bar K}^0_s$ and $B^0_s-{\bar B}^0_s$ mixing and neutron-antineutron oscillations~\cite{Bhaskar1}.

Assuming that $M_N \ll M_X$, one finds an effective four-fermion interaction $N u^c_i d^c_j d^c_k$ after integrating out scalars ${\tilde X}, {\tilde {\bar X}}$. This results in 
decay modes $N \rightarrow p + e^{-} + {\bar \nu}_e , ~ N \rightarrow {\bar p} + e^{+} + \nu_e$, which are kinematically open as long as $M_N > m_p + m_e$ (with $m_p$ and $m_e$ being the proton mass and the electron mass respectively). It is seen that $N$ becomes absolutely stable if $M_N \leq m_p + m_e$. However, in this case, we will have catastrophic proton decay via $p \rightarrow N + e^{+} + \nu_e$ 
if $m_p > M_N + m_e$. 
Therefore a viable scenario with stable $N$ arises provided that
\be \label{massN}
m_p - m_e \leq M_N \leq m_p + m_e \, .
\ee
The important point to emphasize is that stability of $N$ is not related to any new symmetry. It is the stability of the proton, combined with the kinematic condition in Eq.~(\ref{massN}), that ensures $N$ is a stable particle in the above mass window. This leads to a natural realization of GeV DM with or without SUSY~\cite{ad}, which provides a suitable framework to address the DM-baryon coincidence puzzle. Possible signatures of such an ${\cal O}({\rm GeV})$ GeV DM particle in collider and indirect searches have been studied in~\cite{gao1,gao2}.

When $R$-parity is conserved, which we assume to be the case here, the LSP is also a DM candidate. In consequence, if $M_N \approx {\cal O}({\rm GeV})$, a multi-component DM scenario can be realized in this model as both the LSP and $N$ are stable in this case.

After SUSY breaking, the real and imaginary parts of ${\tilde N}$ acquire different masses:
%
\be \label{massNtilde}
m^2_{{\tilde N}_{\rm I,R}} = M^2_N +{\tilde  m}^2 \mp B_N M_N ,
\ee
where ${\tilde m}$ is the soft SUSY breaking mass of ${\tilde N}$ and $B_N M_N$ is $B$-term associated with the $M_N N^2/2$ superpotential term. Depending on the sighn of $B_N$, 
${\tilde N}_{\rm R}$ or ${\tilde N}_{\rm I}$ will be the lighter of the two mass eigenstates. In the special case that $\vert B_N M_N \vert \ll {\tilde m}^2$, ${\tilde N}_{\rm R}$ and ${\tilde N}_{\rm I}$ are approximately degenerate. 
It is clear that there are regions in the parameter space where ${\tilde N}$ (or one of its components) is either the LSP or the NLSP. As we will show below, particularly interesting scenarios can arise with ${\tilde N}$ LSP and gravitino NLSP and vice versa. 


Before closing this section, we briefly comment on the prospects for the detection of $N$ and ${\tilde N}$ as DM candidates in this model. $N$ interacts with nucleons via its coupling to the up qurak that is mediated by the $X$ scalar, see Eq.~(\ref{superpot}). The resulting spin-independent scattering cross section is extremely small as pointed in~\cite{ad}. The spin-dependent scattering cross section is $\sigma^{\rm SD}_{N-{\rm p}} \sim \vert \lambda \vert^2 m^2_{\rm p}/64 \pi m^4_X$~\cite{ad}, where $m_X$ denotes the mass of $X$ scalar. For $m_X \sim {\cal O}({\rm TeV})$ and $\vert \lambda \vert \sim 1$, we have $\sigma^{\rm SD}_{N-{\rm p}} \sim 10^{-42}$ cm$^2$. This is much below the current bounds from direct detection experiments~\cite{coupp}, but within the LHC future reach~\cite{lhc}. The situation is more promising for ${\tilde N}$. It interacts with nucleons via exchange of the $X$ fermion with up quarks, which results in an spin-independent scattering cross section $\sigma^{\rm SI}_{{\tilde N}-{\rm p}} \sim \vert \lambda \vert^2 m^2_{\rm p}/16 \pi M^4_X$~\cite{Rabi2}. For $M_X \sim {\cal O}({\rm TeV})$ and $\vert \lambda \vert^{-1} \leq 10^{-1}$, we have $\sigma^{\rm SI}_{{\tilde N}-{\rm p}} < 10^{-45}$ cm$^2$, which is well within the range currently being probed by direct detection searches~\cite{lux}.

\section{LATE INVISIBLE DECAYS}
\label{sec:invisible}


As mentioned earlier, late decays that involve the gravitino are subject to very tight cosmological constraints from BBN and CMB~\cite{moroibbn,olivebbn,subircmb,hucmb,cobecmb,moroinu}. 
Interestingly, however, the model given in Eq.~(\ref{superpot}) can result in invisible decays to/of the gravitino thereby circumventing these tight cosmological constraints.\footnote{Invisible decays of/to gravitinos involving axino and axion have been discussed in~\cite{steffan}.}  
The two interesting scenarios, pointed out before, are: (1) ${\tilde N}$ NLSP and gravitino LSP, and, (2) ${\tilde N}$ LSP and gravitino NLSP. The decays ${\tilde N} \rightarrow {\tilde G} + N$ (in the former case) and ${\tilde G} \rightarrow {\tilde N} + N$ (in the latter case) do not produce charged particles, hadrons, or neutrinos.\footnote{Secodray production of these particles from the interaction of $N$ and ${\tilde N}$ with nucleons is totally negligible since the corresponding cross section is several orders of magnitude smaller than that of the weak interactions for typical values of the model parameters.}
They are therefore totally invisible and, as a result, evade the above BBN and CMB bounds.
We note, however, that these decays can produce relativistic DM quanta. Cosmological constraints on the model then come from the effective number of neutrinos $N_{\rm eff}$ and from structure formation,
which we will discuss in detail later on.


\begin{figure*}[!t]
\includegraphics[height=4.0cm]{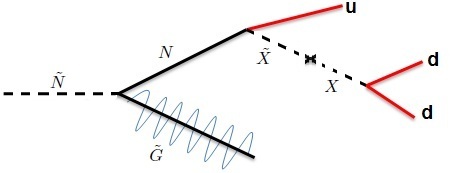}
\caption{
The hadronic decay mode of ${\tilde N}$ in cases 1 and 2. The branching ratio for this channel, given in Eq.~(\ref{branching}), is typically very small and easily satisfies the tightest BBN bounds. A similar diagram exists for hadronic decay of ${\tilde G}$ in case 3 with the role of ${\tilde N}$ and ${\tilde G}$ being reversed.      
}
\label{fourbody}
\end{figure*}

Here we describe different possibilities for a late invisible decay that can arise in our model in more detail:
\\
\\
\noindent
{\bf (1)} ${\tilde N}$ NLSP and gravitino LSP, $m_{\tilde N} > M_N \gg m_{\tilde G}$. The late decay is ${\tilde N} \rightarrow {\tilde G} + N$ with the corresponding width
\be \label{dec1}
\Gamma_{{\tilde N} \rightarrow N + {\tilde G}} = {1 \over 48 \pi} {m^5_{\tilde N} \over M^2_{\rm P} m^2_{\tilde G}} \left(1 - {M^2_N \over m^2_{\tilde N}}\right)^4 .
\ee
For $M_N \approx 1$ GeV, which we consider here, both of the decay products are stable and contribute to the DM relic abundance. However, since $m_{3/2} \ll {\cal O}({\rm GeV})$, the contribution of the gravitino is subdominant. In this case, the dominant component of DM is $N$, while relativistically produced gravitinos from ${\tilde N}$ decay may contribute to the dark radiation.
Gravitinos much lighter than GeV can be realized in models of gauge-mediated SUSY breaking (GMSB). We note that such light gravitinos do not affect stability of $N$ as the only $R$-parity conserving decay mode $N \rightarrow {\tilde G} {\tilde G}$ is forbidden by Lorentz invariance.
\\
\\
\noindent
{\bf (2)} ${\tilde N}$ NLSP and gravitino LSP, $m_{\tilde N} > m_{\tilde G} \gg M_N$. The late decay is ${\tilde N} \rightarrow {\tilde G} + N$ and the corresponding decay width is
\be \label{dec2}
\Gamma_{{\tilde N} \rightarrow N + {\tilde G}} = {1 \over 48 \pi} {m^5_{\tilde N} \over M^2_{\rm P} m^2_{\tilde G}} \left(1 - {m^2_{\tilde G} \over m^2_{\tilde N}}\right)^4 .
\ee
Both of ${\tilde N}$ and $N$ contribute to the DM relic density. However, since $m_{\tilde G} \gg {\cal O}({\rm GeV})$, gravitinos constitute the dominant component of DM, and relativistically produced $N$ quanta from the decay may contribute to the dark radiation.
\\
\\
\noindent
{\bf (3)} ${\tilde G}$ NLSP and ${\tilde N}$ LSP, $m_{\tilde G} > m_{\tilde N} \gg M_N$. The late decay is ${\tilde G} \rightarrow {\tilde N} + N$ and has the following decay width
\be \label{dec3}
\Gamma_{{\tilde G} \rightarrow {\tilde N} + N} = {1 \over 192 \pi} {m^3_{\tilde G} \over M^2_{\rm P}} \left(1 - {m^2_{\tilde N} \over m^2_{\tilde G}}\right)^4 .
\ee
Both of ${\tilde N}$ and $N$ to the DM relic abundance. However, since $m_{\tilde N} \gg {\cal O}({\rm GeV})$, the dominant component of DM is ${\tilde N}$, while relativistically produced $N$ quanta from ${\tilde G}$ decay may contribute to the dark radiation.
\\
\\
Since $M_N \approx 1$ GeV is set by the stability condition of $N$, the parameter space relevant for late decay in all the three cases is two dimensional, namely the $m_{\tilde N}-m_{\tilde G}$ plane. We will discuss in detail the allowed regions of the parameter space for each of these cases later on.

\begin{figure*}[!t]
\mbox{\includegraphics[height=4.0cm]{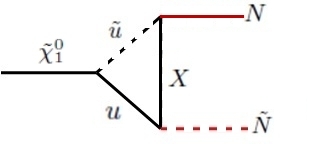} \includegraphics[height=4.0cm]{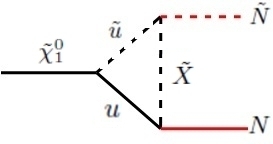}}
\caption{
Typical diagrams for the decay of a Bino-type ${\tilde \chi}^0_1$ into ${\tilde N} N$. Additional diagrams that are obtained by switching $u \leftrightarrow X$ and ${\tilde u} \leftrightarrow {\tilde X}$ in the loop. The SUSY breaking contributions from the diagram on the right dominate giving rise to the decay width in Eq.~(\ref{neutdec}). 
}
\label{loop}
\end{figure*}

Some comments are in order before closing this section. Even though the late decays mentioned above are invisible, they are inevitably accompanied by higher-order processes that produce hadrons and charged particles. One notable channel, shown in Fig. 1, is NLSP decay to three quark final states mediated by an off-shell $N$ and $X,{\bar X}$ scalars. 
The branching ratio for this mode is given by
\be \label{branching}
{\rm Br}_{\rm h} \sim {1 \over (16 \pi^2)^2} \cdot 3 \cdot \vert \lambda \lambda^{\prime}\vert^2 \left({B_X M_X m^2_{\rm NLSP} \over m^4_X}\right)^2 .
\ee
Here $B_X M_X$ is the $B$-term associated with $M_X X {\bar X}$ term in Eq.~(\ref{superpot}), 
and we have assumed $m_{\rm NLSP} \gg m_{\rm LSP}$. 
The first factor on the right-hand (RH) side of Eq.~(\ref{branching}) is the ratio of the phase space factors for four-body and two-body decays respectively, and 3 denotes the color multiplicity factor. 
For $M_X , m_X \gtrsim {\cal O}({\rm TeV})$ (to be compatible with the LHC bounds on the colored particles)
and $\vert B_X \vert \sim {\cal O}({\rm TeV})$, moderately small values of $\vert \lambda \lambda^{\prime} \vert$ and $m_{\rm NLSP} \lesssim 100$ GeV will be enough to push down ${\rm Br}_{\rm h}$ below $10^{-10}$. Such a small value of ${\rm Br}_{\rm h}$ easily satisfies the tight constraints from BBN and CMB bounds mentioned above~\cite{moroibbn,olivebbn,moroinu,subircmb,hucmb}.

Finally, there are other potentially dangerous decay modes that need to be considered. The lightest SUSY particle in the MSSM sector can decay to ${\tilde N}$ and to the gravitino.
These decays can be dangerous if the corresponding lifetime exceeds 1 second. To avoid this, it suffices if the more efficient decay takes place before the onset of BBN. The lightest SUSY particle in the MSSM sector can be the lightest neutralino ${\tilde \chi}^0_1$, the sneutrino ${\tilde \nu}$, or the slepton ${\tilde l}$. 
The neutralino ${\tilde \chi}^0_1$ undergoes two-body decay to ${\tilde N} N$ and four-body decay to $u {\bar u} N {\tilde N}$. The two-body decay occurs via a loop diagram and is dominant. Fig.~2 shows typical diagrams for the decay of a Bino-type ${\tilde \chi}^0_1$. The decay width receives contributions from SUSY preserving and SUSY breaking interactions. The latter dominates by a factor of $(A_X/M_N)^2$, where $A_X$ is the $A$-term associated with the $X N u^c$ superpotential term, see Eq.~(\ref{superpot}). The resulting decay width is:
%
\be \label{neutdec}
\Gamma_{{\tilde \chi}^0_1} \sim {4 \over 3} \cdot {1 \over (4 \pi)^4} \cdot {1 \over 8 \pi} \alpha \vert \lambda \vert^4 {m^3_{{\tilde \chi}^0_1} A^2_X \over M^4_X} .
\ee
Here $\alpha$ is the electroweak fine structure constant,factors of $4/3$ and $1/(4 \pi)^4$ on the RH side take the the color multiplicity and hypercharge of the up-type quarks and the loop factor into account respectively, and we have assumed $m_{{\tilde \chi}^0_1} \gg m_{\tilde N}$. 
As mentioned before, only the combination $\vert \lambda \lambda^{\prime} \vert$ is subject to phenomenological constraints in our model, and hence $\vert \lambda \vert$ does not need to be very small. For $A_X,M_X \sim 1$ TeV, $m_{{\tilde \chi}^0_1} \sim 100$ GeV, and $\vert \lambda \vert \sim 10^{-2}$, we find $\tau_{{\tilde \chi}^0_1} \ll 10^{-6}$ sec. Combination of SUSY breaking and electroweak breaking interactions lead to similar decay widths for Wino-type and Higgsino-type ${\tilde \chi}^0_1$ via one-loop diagrams. The sneutrino ${\tilde \nu}$ and slepton ${\tilde l}$ decay to $\nu N {\tilde N}$ and $l N {\tilde N}$ final states via an off-shell Bino, and the corresponding decay widths are still $\ll 1$ sec. We therefore see that for reasonable choice of parameters the lightest SUSY particle in the MSSM sector decays early enough to easily avoid any potential danger.

\section{DARK RADIATION Constraints on LATE Invisible DECAYS}
\label{sec:derive}
The amount of dark radiation in the early universe is parametrized by the effective number of neutrinos $N_{\rm eff}$. The present observational bound on $\Delta N_{\rm eff} \equiv N_{\rm eff}-N_{\rm eff, SM}$ (where $N_{\rm eff} = 3.04$) from Planck+WMAP9+ACT+SPT+BAO+HST at 2$\sigma$ is $\Delta N_{\rm eff} = 0.48^{+0.48}_{-0.45}$~\cite{planck}, which implies that $\Delta N_{\rm eff} = 0.96$ at 2$\sigma$. The value precise of $N_{\rm eff}$ depends on Hubble constant where the Planck data and HST measurements differ~\cite{hst}. The reconciliation can occur using a non-zero $\Delta N_{\rm eff}$~\cite{Wyman:2013lza}. Setting aside the dust contamination in the BICEP2 results, the tension in the CMB tensor polarization measurement between the recent BICEP2~\cite{bicep2} and Planck data can also be reconciled with $\Delta N_{\rm eff}$=0.81$\pm$0.25 at more than $3 \sigma$ confidence level in a joint analysis~\cite{Dvorkin:2014lea}, disfavoring $\Delta N_{\rm eff}=0$.  Since the presence of dark radiation is debatable, here we take a more conservative approach. We use the data the derive bounds on frameworks that naturally may induce a non-negligible dark radiation component through non-thermal DM production.

In order to relate the energy density associated with non-thermally, relativistically produced DM with the effective number of neutrinos, we start by calculating the ratio between their respective energy densities. Since the cold dark matter (CDM) and neutrino energy densities are redshifted like $\rho_{\rm DM} \propto \Omega_{\rm DM} a^{-3}$ and $\rho_{\nu} \propto \Omega_{\nu} a_{eq}^{-4} N_{\nu}/3$, 
the ratio between the neutrino and DM energy densities at the matter-radiation equality is
\begin{equation}
\frac{\rho_{\nu}}{\rho_{\rm DM}}= \frac{\Omega_{\nu}}{\Omega_{\rm DM}}\ \frac{N_{\nu}}{3}\ \frac{1}{a_{\rm eq}}=\frac{0.69\ \Omega_{\gamma}}{\Omega_{\rm DM}} \frac{N_{\nu}}{3} \frac{1}{a_{\rm eq}},
\label{rhoDMnu}
\end{equation}
where $\Omega_{\gamma} \simeq 4.84 \times 10^{-5}$, $\Omega_{\rm DM} \sim 0.227$, $N_{\nu}$ is the number of neutrinos.
For $N_{\nu}=1$, we thus find that the energy density of one neutrino is $\sim16\%$ of the DM density. As a result, if DM particles had a kinetic energy equivalent to $\gamma_{\rm DM} \simeq 1.16$ at $t_{\rm eq}$ 
, this fraction would produce the same effect as an extra neutrino species in the expansion of the universe at that time
~\cite{Hooper:2011aj}. However, as we will discuss below, constraints stemming from structure formation require the fraction of DM particles with appreciable kinetic energy to be $\ll 1$. Therefore, in order to still mimic one neutrino species a small fraction of DM particles have to be relativistically produced. 

\begin{figure*}[!t]
\includegraphics[height=8.0cm]{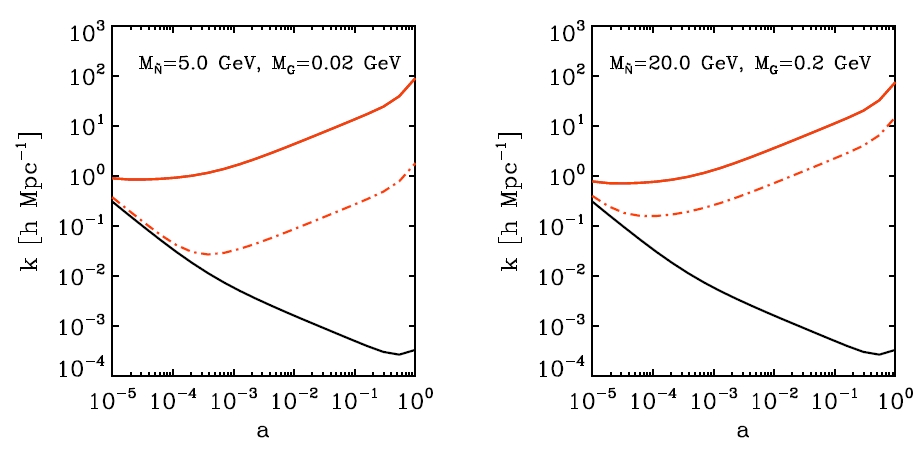}
\caption{
The free-streaming scale $k_{\rm fs}$ for N $\&$ $ \tilde{G}$ for the decay process $\tilde{N} \rightarrow \tilde{G} + N$ (case 1) as a function of scale factor a. Each panel is for different value of $m_{\tilde{G}}$ and $m_{\tilde{N}}$. For the left panel the mass is set to $m_{\tilde{N}}$ = 5 GeV and $m_{\tilde{G}}$ = 0.02 GeV; in the right panel masses are set to $m_{\tilde{N}}$ = 20 GeV and $m_{\tilde{G}}$ = 0.2 GeV. The mass of $m_N$ is 1 GeV. The black solid line is the size of the horizon at a given scale factor. The solid color lines are for N, and dash-dotted lines are for $\tilde{G}$. The density perturbation is suppressed for k $>$ $k_{\rm fs}$. Between $k_{\rm fs}$ and the horizon the density perturbation will grow.
}
\label{fig:m2}
\end{figure*}

In the general decay setup where a heavy particle with mass $M$ decays at rest to two particles with masses $m_1$ and $m_2$ at time $t_{\rm dec}$, the boost factors of the daughter particles follows
%
%
\begin{eqnarray}
\gamma_{1}(t)^2 & = & 1+\frac{a^2_{\rm dec}}{a^2(t)}\frac{p^2}{m_{1}^2} \, , \nonumber \\
\gamma_{2}(t)^2 & = & 1+\frac{a^2_{\rm dec}}{a^2(t)}\frac{p^2}{m_{2}^2} \, .
\label{boost}
\end{eqnarray}
where
%
%
\begin{eqnarray}
p = \frac{\left[\left(M^2 - (m_{1}+m_{2})^2\right)
\left(M^2 - (m_{1}-m_{2})^2\right)\right]^{1/2}}{2 M} \, , \nonumber \\
\,
\label{moment}
\end{eqnarray}
is the momentum of the daughter particles at the time of production.
%
%

If both of the daughter particles are stable, they both contribute to the DM relic density. If $f$ is the fraction of the energy density in DM that is produced from the late decay, the amount of dark radiation that is mimicked by the kinetic energy of the daughter particles is found to be
\be \label{neff1}
\Delta N_{\rm eff} = \left[{(\gamma_{1,{\rm eq}} - 1) m_1 + (\gamma_{2,{\rm eq}} -1) m_2 \over 0.16 (m_1 + m_2)}\right] f.
\ee
%
%
We note that the normalization factor 0.16 above appears due to the neutrino-DM energy density fraction at the matter radiation equality according to Eq.~(\ref{rhoDMnu}).

If $m_1 \ll m_2$, then the likely scenario is that species 1 is the dominant DM component, hence $m_{\rm DM} \simeq m_1$, while species 2 makes the major contribution to the dark radiation.
\footnote{The case when the same species is responsible for DM and dark radiation has been studied in detail in~\cite{Hooper:2011aj,Kelso:2013paa,Kelso:2013nwa,Queiroz:2013lca}.} In this case, assuming that $\gamma_{2,{\rm eq}} \gg 1$, we have
\be \label{neff2}
\Delta N_{\rm eff} \simeq 4.87 \times 10^{-3} \left(\frac{t_{\rm dec}}{10^6\ s} \right)^{1/2} \left(\frac{p}{m_{\rm DM}}\right) f ,
\ee
where $p$ is given in Eq.~(\ref{moment}).

%
%
%

\section{Structure formation constraints on Late Invisible Decays} 
\label{sec:structure}
In this section, we discuss large scale structure constraints and present our results for dark radiation in the model described in section II. 
The median speed of the decay products at a given time $t$ for $M \rightarrow m_1 + m_2$ is described by
\beq
v_{1,2,{\rm med}}(t) \sim {a_{\rm dec}/a(t)p \over \sqrt{(a_{\rm dec}/a(t)p)^2+m_{1,2}^2}}
\eeq
%


The scaling of the free-streaming distance can be understood in terms of the Jeans wavenumber:
\beq
k_{\rm fs}(a)={\sqrt{\rho a^2/2 M^2_{\rm P}} \over v_{\rm med}(a)} = \sqrt{{3\over2} }{aH(a) \over v_{\rm med}(a)}, 
\eeq
where for k $>$ $k_{\rm fs}$, the density perturbation is suppressed.


Correlation of the galaxy distribution probes the matter power spectrum on scales of 0.02 h $\mathrm{Mpc}^{-1}$ $\lsim$ k $\lsim$ 0.2 h $\mathrm{Mpc}^{-1}$ at z $\sim$ 0~\cite{Reid_etal10}. Indeed one of the best cosmological probes of constraining massive standard model neutrinos, as a class of ``hot dark matter'' (HDM), is galaxy power spectrum. The current neutrino mass limits from SDSS galaxy clustering is about $\Sigma m_{\nu} <$ 0.3-0.62 eV~\cite{Reid_etal10}. The abundance of HDM that is allowed by current galaxy power spectrum is given by 
\beq
\Omega_{\nu}h^2 = {\Sigma m_{\nu} \over 94.1 \mathrm{eV}}. 
\eeq
This predicts $\Omega_{\nu} \lsim$ 0.007-0.01, which gives the ratio of ``DM'' and ``HDM'' $\Omega_{\nu}/\Omega_{\rm DM} \lsim$ 0.03-0.06. Therefore the amount of ``HDM'' that suppresses structure growth at scale $k \sim 0.02 h \mathrm{Mpc}^{-1}- 0.2 h \mathrm{Mpc}^{-1}$, either from the subdominant part or the dominant part of the dark matter, is limited to be less than 3-6$\%$ of the total dark matter.

Lyman-alpha forest data probes the matter power spectrum on smaller scales, 0.1 h $\mathrm{Mpc}^{-1}$ $\lsim$ k $\lsim$ 2 h $\mathrm{Mpc}^{-1}$ at z $\sim$ 2$-$4 \cite{McDonald_etal06, Palanque-Delabrouille_etal13}. For current Lyman-alpha forest data, the error of measurement is roughly in the range of 5-10$\%$~\cite{McDonald_etal06, Palanque-Delabrouille_etal13}. In~\cite{boyarsky_etal08}. These authors utilize numerical simulations to study Lyman-alpha forest limits for the warm+cold dark matter models. They found that, with fraction of sterile neutrino warm dark matter (WDM) $f_{\rm WDM} <$ 0.35 any mass of WDM in the range they studied is allowed by the data. So the amount of ``WDM'' which suppresses structure growth at scale k $\sim$ 0.1 h $\mathrm{Mpc}^{-1}$ $-$ 2 h $\mathrm{Mpc}^{-1}$ cannot be more than 10$-$35$\%$ of the total dark matter.

\begin{figure*}[!htp]
\includegraphics[scale=2.2]{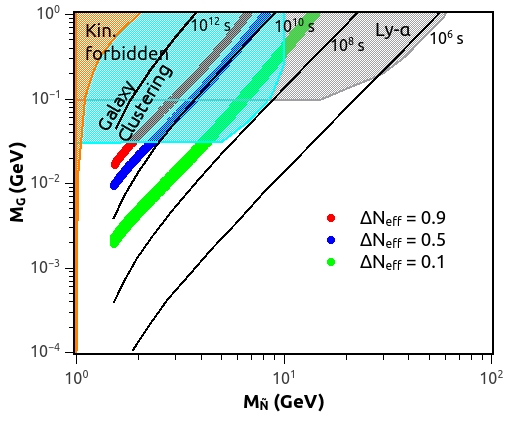}
\caption{
The allowed region of the parameter space in case 1 (${\tilde N} \rightarrow {\tilde G} + N$, $m_{\tilde G} \ll M_N \approx 1$ GeV) is shown. 
Lifetime contours of ${\tilde N}$ and $\Delta N_{\rm eff} = {\rm const}$ bands are included. }
\label{fig1}
\end{figure*}

In Fig.~(\ref{fig:m2}), we plot the free-streaming scale $k_{fs}$ for $N$ and $ \tilde{G}$ for the decay process $\tilde{N} \rightarrow \tilde{G} + N$ (case 1) as a function of scale factor $a$. Each panel is for different value of $m_{\tilde{G}}$ and $m_{\tilde{N}}$. We find that $k_{\rm fs}$ for the decay products, depending on the masses, can suppress scales that can be probed by the large scale structure data.

We note that constraints on our late invisible decay models from estimates of the free-streaming length are meant to provide rough estimates. To understand the power spectrum suppression scale accurately, one will need to solve the Boltzmann equations and derive the perturbation evolution~\cite{Kaplinghat:2005sy,Strigari:2006jf}. However, this is beyond the scope of this study, and we will leave this for future work.

\begin{figure*}[!htp]
\includegraphics[scale=2]{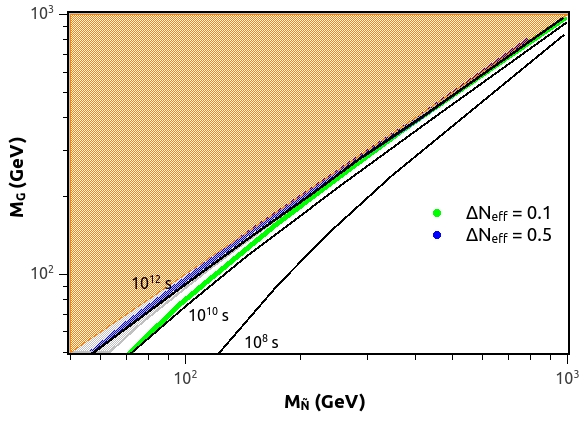}
\caption{
The same as Fig. 4 but for case 2 (${\tilde N} \rightarrow {\tilde G} + N$, $m_{\tilde G} \gg M_N \approx 1$ GeV). 
}
\label{fig2}
\end{figure*}

\begin{figure*}[!htp]
\mbox{\includegraphics[scale=2]{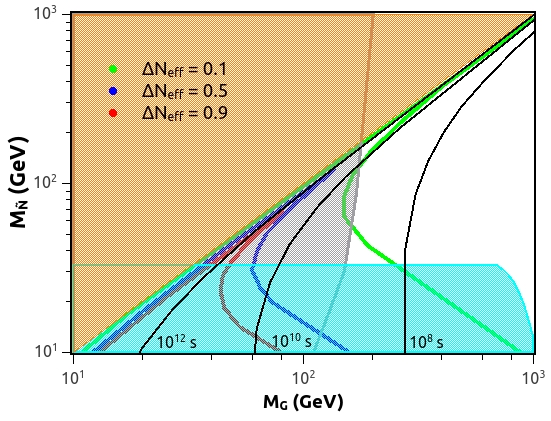}}
\caption{The same as Fig. 4 but 
for case 3 (${\tilde G} \rightarrow {\tilde N} + N$, $m_{\tilde N} \gg M_N \approx 1$ GeV). 
}
\label{fig3}
\end{figure*}

\section{Results}
\label{sec:result}
In this section, we present our results.
In figures~\ref{fig1}-\ref{fig3}
we include the constraints from structure formation and plot $\Delta N_{\rm eff}$
in the $m_{\rm NLSP}-m_{\rm LSP}$ parameter space. 
The figures depict contours for the decay lifetime $t_{\rm dec}$ of the NLSP and bands representing the value of $\Delta N_{\rm eff}$. We have shown  LSP masses up to 1 TeV ; for larger masses the SUSY particles will be too heavy to have a realistic prospect for their detection at the LHC. We also note that the region $m_{\rm NLSP} \leq m_{\rm LSP} + M_N$, where $M_N \approx 1$ GeV, is kinematically forbidden.

Fig.~(\ref{fig1}) shows the results for case 1, ${\tilde N} \rightarrow N + {\tilde G}$ decay with $m_{\tilde N} > M_N \gg m_{\tilde G}$.
In this case $N$ is the dominant component of DM, and gravitino quanta from ${\tilde N}$ decay make the main contribution to dark radiation. The corresponding decay width is given by Eq.~(\ref{dec1}) where $M_N \approx 1$ GeV. Since the decay creates the same number of $N$ and ${\tilde G}$ quanta, the contribution of gravitinos to the total DM density is $f m_{\tilde G}/1~ {\rm GeV}$, where $f$ is the ratio of $N$ number density from ${\tilde N}$ decay to its total value.
We take $f=1$ henceforth, which results in the tightest bounds from structure formation and ${\Delta N}_{\rm eff}$.
If we use smaller values of $f$, the constraints will become weaker, but we also need to have additional source of DM.

%
%
%
For $m_{\tilde N} \gg 1$ GeV, Eqs.~(\ref{dec1},\ref{moment},\ref{neff2})
result in ${\Delta N}_{\rm eff} \propto m_{\tilde G}/m^{3/2}_{\tilde N}$, implying that ${\Delta N}_{\rm eff} = {\rm const}$ bands lie along the curves $m^{3}_{\tilde N} \propto {\Delta N}^2_{\rm eff} m^2_{\tilde G}$. We note, however, that $m_{\tilde G}$ eventually catches up with $m_{\tilde N}$, at which point the decay becomes kinematically impossible. Therefore, the ${\Delta N}_{\rm eff} = {\rm const}$ bands have a turning point where they bend to the left as seen in the figure.

We find that the Lyman-alpha forest data is effective in constraining the parameter space for $m_{\tilde N}>$10 GeV, while the galaxy power spectrum sets a stronger constraint for $m_{\tilde N}<$10 GeV. In the latter case, the gravitino mass needs to be smaller than 0.01 GeV. We see that it is still possible to get $\Delta N_{\rm eff}\sim$ 0.5 in the allowed region of the parameter space for $m_{\tilde N}<$2 GeV.

Fig.~(\ref{fig2}) shows the results for case 2, ${\tilde N} \rightarrow N + {\tilde G}$ decay with $m_{\tilde N} > m_{\tilde G} \gg M_N$.
In this case gravitino is the dominant component of DM, and $N$ quanta from ${\tilde N}$ decay make the main contribution to dark radiation. The corresponding decay width is given by Eq.~(\ref{dec2}) where $M_N \approx 1$ GeV.

Eqs.~(\ref{dec2},\ref{moment},\ref{neff2}) result in
%
%
%
$\Delta N_{\rm eff} \propto m^{1/2}/(m^2_{\tilde N} - m^2_{\tilde G})$. This implies that ${\Delta N}_{\rm eff} = {\rm const}$ bands are concentrated around the line $m_{\tilde N} = m_{\tilde G}$ for large values of $m_{\tilde N}$ as seen in the figure.

We find that the Lyman-alpha forest data is effective in constraining the parameter space for $m_{\tilde{G}}>$100 GeV, while the galaxy power spectrum sets a stronger constraint for $m_{\tilde{G}}<$100 GeV. Combining both constraints, we find that $\Delta N_{\rm eff} \lesssim 0.1$ in the allowed region of the parameter space.

Fig.~(\ref{fig3}) shows the results for case 3, ${\tilde G} \rightarrow {\tilde N} + N$ decay with $m_{\tilde G} > m_{\tilde N} \gg M_N$.
In this case ${\tilde N}$ is the dominant component of DM, and $N$ quanta from gravitino decay make the main contribution to dark radiation. The corresponding decay width is given by Eq.~(\ref{dec3}) where $M_N \approx 1$ GeV.

For $m_{\tilde G} \gg m_{\tilde N}$ Eqs.~(\ref{dec3},\ref{moment},\ref{neff2}) result in
%
%
${\Delta N}_{\rm eff} \propto m^{1/2}_{\tilde G}/m_{\tilde N}$. Therefore ${\Delta N}_{\rm eff} = {\rm const}$ bands lie along the curves $m^2_{\tilde N} \propto {\Delta N}^{-2}_{\rm eff}/m_{\tilde G}$. 
Along this curve $m_{\tilde N}/m_{\tilde G}$ decreases as $m_{\tilde G}$ becomes smaller. 
Hence, in order for the decay to be kinematically allowed, the ${\Delta N}_{\rm eff} = {\rm const}$ bands should eventually bend to the right. On the other hand, when $m_{\tilde N} \simeq m_{\tilde G}$, 
we have ${\Delta N}_{\rm eff} \propto 1/m^{1/2}_{\tilde G} (m_{\tilde G} - m_{\tilde N})$. This implies that the ${\Delta N}_{\rm eff} = {\rm const}$ bands join together around the $m_{\tilde N} = m_{\tilde G}$ line after turning to the right as seen in the figure.

We find that the Lyman-alpha forest data is effective in constraining the parameter space for $m_{\tilde{G}}<$150 GeV, while the galaxy power spectrum sets a stronger constraint for $m_{\tilde{N}}<$100 GeV. Combining both constraints, we find that $\Delta N_{\rm eff} \lesssim 0.5$ in the allowed region of the parameter space.

\section{CONCLUSION}
\label{sec:conclusion}
In this work, we investigated a simple extension of the MSSM that accommodates 
late invisible decays to/of the gravitino. 
The model includes new iso-singlet color-triplet superfields $X$ and ${\bar X}$ and a singlet superfield $N$. Such extension allows us to explain the baryon asymmetry of the universe, and can also address the DM-baryon coincidence puzzle. 
In addition to the LSP, this model has an $R$-parity even DM candidate when the singlet fermion $N$ has ${\cal O}({\rm GeV})$ mass. 

Interesting cases arise when the singlet scalar ${\tilde N}$ and the gravitino ${\tilde G}$ are the NLSP and LSP, respectively, and vice versa. The resulting decays ${\tilde N} \rightarrow {\tilde G} + N$ and ${\tilde N} \rightarrow {\tilde G} + N$ have long lifetimes as they involve gravitationally suppressed interactions. 
However, since both of the outgoing particles are invisible, 
these late decay
are not subject to the tight BBN and CMB constraints on the hadronic and electromagnetic channels. 
On the other hand, depending on the mass ratios of the daughter and parent particles, it is possible that one or both of the DM candidates 
contribute to the amount of dark radiation 
or suppress perturbations at scales that are being probed by the galaxy cpower spectrum and the Lyman-alpha forest data. 
We performed a detailed study of the $m_{\tilde N}-m_{\tilde G}$ parameter space in light of these constraints 
and showed that the entire DM content of the universe can be
produced from the late invisible decays to/of the gravitino. 
Such decays have very important consequences in a broader context as they considerably relax the constraints on reheating of the universe in SUSY models.
\vskip 1mm
\acknowledgments
We would like to thank Yu Gao for helpful discussions. The work of R.A. is supported in part by NSF Grant No. PHY-1417510. The work of B.D. is supported by DOE Grant DE-FG02-13ER42020. The work of L.E.S. and M.W is supported in part by NSF Grant No. PHY-1417457. F.Q. is partly supported by DOE Award SC0010107 and the Brazilian National Counsel for Technological and Scientific Development (CNPq).


\end{document}